\documentclass[galaxies,article,submit,pdftex,moreauthors]{Definitions/mdpi} 

\firstpage{1} 
\pubvolume{1}
\issuenum{1}
\articlenumber{0}
\pubyear{2024}
\copyrightyear{2024}
\datereceived{ } 
\daterevised{ } 
\dateaccepted{ } 
\datepublished{ } 
\hreflink{https://doi.org/} 

\usepackage{graphicx} 
\usepackage{amsmath}
\usepackage{amssymb}
\usepackage{caption}
\usepackage{comment}
\usepackage{subcaption}
\usepackage{geometry}
\usepackage{wrapfig}
\usepackage{enumitem}
\usepackage{mathtools}
\usepackage{matlab-prettifier}
\usepackage{bm}
\usepackage{svg}
\usepackage{ltablex}

\newlength\figurewidth

\Title{Non-Smooth Multi-objective Controller Synthesis for Test-Mass Actuation in Gravitational-Wave Detectors}

\TitleCitation{Non-Smooth Multi-objective Controller Synthesis for Test-Mass Actuation in Gravitational-Wave Detectors}

\Author{Sander K. Sijtsma $^{1,\dagger,*\orcidA{}}$, 
Pooya Saffarieh $^{2, 3}$\orcidD{}, 
Nathan A. Holland $^{2, 3}$\orcidB{}, 
Sil T. Spanjer $^{1}$\orcidE{},
Wouter B. J. Hakvoort $^{1}$\orcidF{},
and Conor M. Mow-Lowry $^{2, 3}$\orcidC{}}

\AuthorNames{Sander K. Sijtsma, Pooya Saffarieh, Nathan A. Holland, Sil T. Spanjer, Wouter B. J. Hakvoort and Conor M. Mow-Lowry}

\AuthorCitation{Sijtsma, S.K.; Saffarieh, P.; Holland, N.; Spanjer, S.T.; Hakvoort, W.B.J; Mow-Lowry, C.M.}

\address{%
$^{1}$ \quad Department of Mechanics, Solids, Surfaces and Systems, University of Twente,7522 NB Enschede, Netherlands\\
$^{2}$ \quad Nikhef, 1098 XG Amsterdam, Netherlands\\
$^{3}$ \quad Department of Physics and Astronomy, Vrije Universiteit Amsterdam, 1081 HV Amsterdam, Netherlands
}

\corres{Correspondence: ssijtsma@nikhef.nl;}

\firstnote{Currently employed at Nikhef}  

\nolinenumbers
\abstract{
This paper proposes a non-smooth controller optimization method and shows the results of ongoing research on the implementation of this method for gravitational wave applications. Typical performance requirements concerning these type of suspensions are defined in terms of both $\mathcal{H}_2$- and $\mathcal{H}_{\infty}$-type constraints. A non-smooth optimization approach is investigated, which allows the use of non-convex cost functions that are often a result of mixed $\mathcal{H}_2/\mathcal{H}_{\infty}$ optimization problems. Besides the controller, the distribution of the actuation is integrated with the optimization to investigate the feasibility of simultaneous controller and actuator optimization. The results demonstrate that the proposed non-smooth optimization method is able to find suitable solutions for the control and actuator distribution that satisfy all required performance and design constraints.}


\begin{document}


\section{Introduction}

Observations of gravitational waves made by the LIGO-Virgo-KAGRA collaboration \cite{GWTC1,GWTC2,GWTC3,GWTC4,GWTC5,GWTC6,GWTC7} are the result of decades of innovation of the ground-based detectors LIGO \cite{AdLigo}, Virgo \cite{AdVirgo}, and KAGRA \cite{Kagra}. All of these detectors employ laser-interferometry to detect gravitational waves that pass through the earth. The mirrors, or test masses, that reflect the laser at the ends of the interferometer arms are isolated from vibration by means of extensive suspension systems that include both active and passive vibration isolation strategies \cite{seismic_isolation,aLIGOquad}. Although current and future detectors have adapted different vibration isolation techniques, each observatory's vibration isolation involves a `payload' suspension  that refers to the final stages of the vibration isolation. Actuation on the payload is the primary method for compensating residual disturbances and `locking' the interferometer. The global feedback controllers, required to lock the optical resonators, are usually designed via classical loop-shaping methods. While these methods are effective, they tend to be challenging for less experienced control designers, and it is time-consuming to develop many such controllers during the design phase of the suspension system and payload.

More modern approaches for the design of control systems are automated controller optimization strategies, amongst which $\mathcal{H}_2$- and $\mathcal{H}_{\infty}$-synthesis are among the most commonly employed methods \cite{ref-feedbackcontrol}, and they have already been applied for gravitational-wave applications \cite{hinf_sensor_fusion,mixedVirgo}. Typically, the computation of a suitable control algorithm that meets the control system requirements is performed by a computer program, and the responsibility of the control system designer is shifted to translating the design requirements into a relevant optimization problem. The benefit of this approach lies in the fact that once a suitable mathematical definition of the control problem exists, a change in the noise models or the observatory's requirements is reflected simply by the relevant parameters within the problem definition. The computational program produces an updated controller design without much additional effort. Compared with classical controller design methods, this optimal approach can simplify and accelerate the design or redesign of suitable control algorithms. Moreover, optimization methods allow for quick performance evaluation for a wide range of suspension configurations during the conceptual design phase enabling a more holistic controls and mechanics design process. As such, optimal control methods are especially interesting for third generation gravitational-wave detectors the Einstein Telescope (ET) and Cosmic Explorer that are still in the development phase \cite{et_report, ce_report}. Specifically, the sensitivity design curve and requirements for the low-frequency optimized interferometer of the Einstein Telescope, ET-LF, are considered in this study.

This paper investigates the suitability of a non-smooth mixed synthesis optimization algorithm  \cite{apkarian-2009, apkarian-2005}, which allows us to automatically tune controllers for a problem that is constrained by both $\mathcal{H}_2$ and $\mathcal{H}_{\infty}$ system norms. This approach was successfully applied to similar applications within the precision engineering field \cite{ref-aspe}.  A simplified model of the suspension is considered to illustrate how this optimization approach fits the actuation requirements typical for a payload suspension gravitational-wave detectors \cite{losurdo_2001}. Moreover, the optimization of the distribution of the actuator forces is considered, opening the possibility to jointly design both the controller and a parametric mechatronic design.


\section{Modelling}
\label{ch:modelling}
\begin{wrapfigure}{r}{0.3\textwidth}
    \centering
    \includegraphics[width=3.5cm]{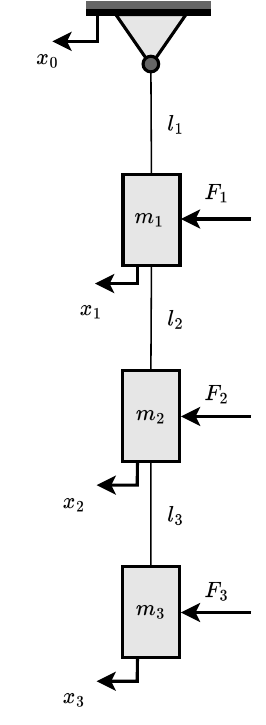}
    \caption{Schematic model of the payload suspension}
    \label{fig:IPM_payload}
\end{wrapfigure}

In order to demonstrate non-smooth optimisation methods, the dynamics of the payload suspension can be represented with sufficient accuracy by a multi-pendulum system. In this study, we consider a 3-stage, three Degree of Freedom (DoF) pendulum that is actuated at the mirror and both upper stages. A schematic model of the plant is shown in Figure \ref{fig:IPM_payload}. The mass of each stage is denoted as $m_{\rm{i}}, \hspace{2mm} i = 1,2,3$ and the length of each pendulum by $l_{\rm{i}}, \hspace{2mm}i=1,2,3.$ The values for masses and lengths are taken from \cite{koroveshi_23}.

The residual seismic disturbance at the base of the payload suspension is denoted by the signal $x_0(t)$. The state vector of the suspension is given by the angles of the pendulums as $\bm{q} = [\theta_1\hspace{1mm} \theta_2\hspace{1mm} \theta_3]^T$. Although the pendulum angles are used to express the equations of motion that describe the dynamics of the suspension, the measured coordinate that we are actually interested in is the displacement of the mirror. When the equations of motion are converted into state-space representation, the $C$-matrix converts angles into displacement such that the output of the model is $y \approx l_1\theta_1 + l_2\theta_2 + l_3\theta_3$. The aim of the suspension system is to minimize the coupling of $x_0(t)$ to the motion $x(t)$ of the mirror. The general form of the linearized dynamics of the payload suspension is given by the following equation, assuming the small angle approximation
\begin{equation}
    M\bm{\ddot{q}} + g\bm{q} = \bm{\xi},
\end{equation}
where
\begin{align}
    M =  \begin{bmatrix*}[c]
        (m_1+m_2+m_3)l_1^2 & (m_2+m_3)l_1l_2 & m_3l_1l_3 \\
        (m_2+m_3)l_1l_2    & (m_2+m_3)l_2^2  & m_3l_2l_3 \\
        m_3l_1l_3          & m_3l_2l_3       & m_3l_3^2
     \end{bmatrix*},
\end{align}

\begin{align}
    g = \begin{bmatrix*}
         (m_1+m_2+m_3)gl_1  & 0                     & 0         \\
         0                  & (m_2+m_3)gl_2         & 0         \\
         0                  & 0                     & m_3gl_3
     \end{bmatrix*}.
\end{align}

Since this plant has multiple inputs and a single measured output, the system is of Multiple Input Single Output (MISO) type. 

\section{Problem Formulation}

The motion of the mirror is interferometrically sensed by the observatory's primary laser through the `global' interferometer sensing and control system. The main objective for the global interferometer control system is to `lock' the interferometer by keeping it within its small linear operating range. This is necessary to achieve the extreme sensitivities that enable the detection of gravitational waves. A driving requirement of this control system is to limit the differential motion of the optical resonators within the detector arms. This motion must be reduced to less than a picometer \cite{aLIGOquad}. We assume a  value of $1\cdot10^{-13}$\,m in this work. The dominant contribution to this differential motion is the residual seismic motion leaking through the active and passive isolation stages at low frequencies, typically between 0.02\,Hz and 0.2\,Hz.

Control of this motion is distributed over the final stages of the mirrors' suspension. This necessarily requires actuators that introduce (Digital to Analog Converter) DAC noise to the system, since the control system is implemented digitally. Since the sensitivity of a gravitational wave detector is often expressed as a combination of the open-loop equivalent contributions from several sources, the contribution of the open-loop equivalent DAC noise may not exceed the observatory's design sensitivity curve at any point over the entire sensitive frequency range. Tables of all the signals and dynamic systems that will appear in the following paragraphs are given in Appendix \ref{app:app1} and \ref{app:app2}.

\begin{figure}[H]
    \centering
    \includegraphics[width=12.95cm]{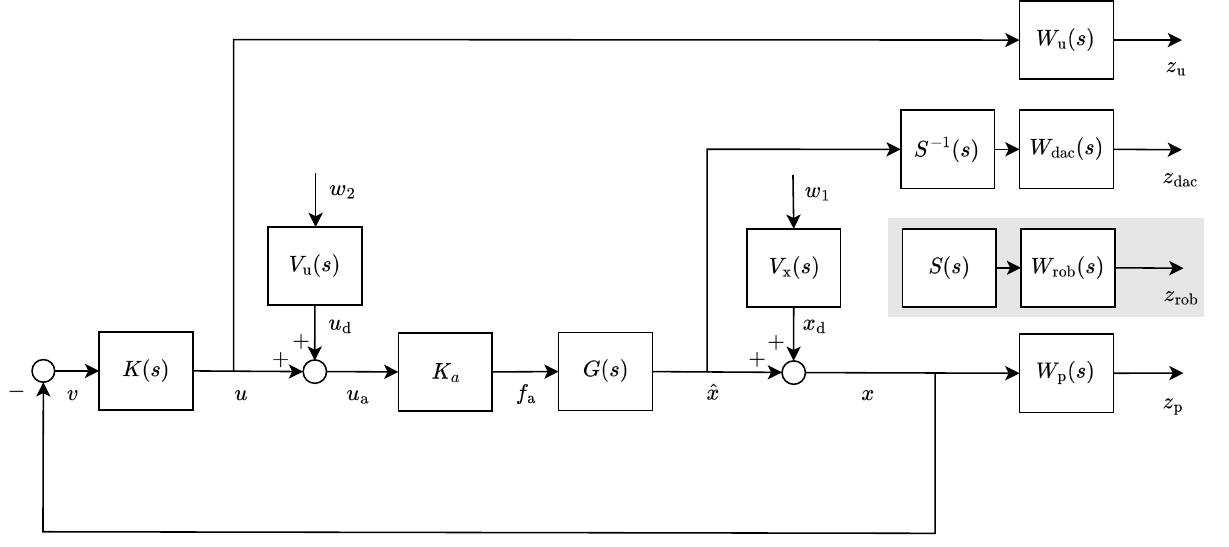}
    \caption{Block diagram of the payload suspension control system, including $\mathcal{H}_2/\mathcal{H}_{\infty}$-weights}
    \label{fig:block_diagram}
\end{figure}

The main performance objective of the control system is to find a controller and actuator distribution that limits the closed-loop root mean square (RMS) of the mirror's residual motion. This is a typical $\mathcal{H}_2$-control problem, since an $\mathcal{H}_2$-optimal controller effectively minimizes the variance of selected closed-loop control signals. The requirement on the DAC noise is a hard limit. The open-loop equivalent spectra must not exceed the detector's sensitivity curve. This can be captured by an $\mathcal{H}_{\infty}$-norm on the open-loop equivalent DAC noise spectrum. Moreover, desired robustness margins typically manifest as additional $\mathcal{H}_{\infty}$-constraints. Figure \ref{fig:block_diagram} shows a block diagram of the payload suspension control system.

The transfer function $G(s)$ is a model of the suspension that maps actuator forces to the mirror motion. The dynamics of the actuators are represented by a frequency-independent matrix $K_{\rm a}$\,N/V such that the actuators can be simultaneously optimized together with the controller to reduce the coupling of DAC noise into the system. Choosing a flat gain to represent the actuator dynamics simplifies the optimization problem, since we are neglecting high-frequency dynamics that are not interesting for this problem. The gain represents the size of the actuator, such that the optimizer can return an optimal distribution of the actuation over the three stages of the suspension. The total transfer from input voltage $u_{\rm{a}}$ to displacement of the mirror $\hat{x}$ is then given by $G(s)K_{\rm a}$. The signal $x$ denotes the total mirror motion, including the residual seismic disturbance, that is present at the mirror. Channels for both relevant $\mathcal{H}_2$ and $\mathcal{H}_{\infty}$ constraints are also included in the block diagram.

An additional $\mathcal{H}_2$-constraint $W_{\rm{u}}(s)$ on the control signal $u$ is introduced to prevent the RMS of this signal from exceeding the DAC range. The RMS requirement is captured by the weight $W_{\rm{p}}$, which is a constant gain. 
The hard limit on the open-loop-equivalent DAC noise is dictated by the filter $W_{\rm dac}(s)$, which modeled as the approximate inverse of the sensitivity curve of the detector \cite{gw_moore}.

Notice that the inverse of the sensitivity function $S(s)$ is introduced to emphasize that the DAC noise optimization channel is related to an open-loop requirement. An $\mathcal{H}_2$-optimal controller often results in a closed-loop system with very small stability margins, which is often undesirable for any control system. A robustness margin can be defined as the distance between the critical point and the loop gain in the Nyquist plot. This robustness margin is guaranteed by introducing an additional $\mathcal{H}_{\infty}$-constraint $W_{\rm{rob}}$ that is applied to the sensitivity function to maintain a desired distance between the loop gain and the critical point on the Nyquist plot. The sensitivity is the closed-loop transfer function that maps the seismic disturbance $x_{\rm{d}}$ to the mirror motion $x$ and is defined as

\begin{equation}
    S(s) =  (I+G(s)K_{\rm{a}}K(s))^{-1}.
\end{equation}

Peaking of this sensitivity function is constrained with an $\mathcal{H}_{\infty}$-bound, since the robustness margin of the system is inversely proportional to the $\lVert S(s) \rVert_{\infty}$-norm.

\section{Optimization}

\begin{figure}[H]
    \centering
    \includegraphics[width = 8.1cm]{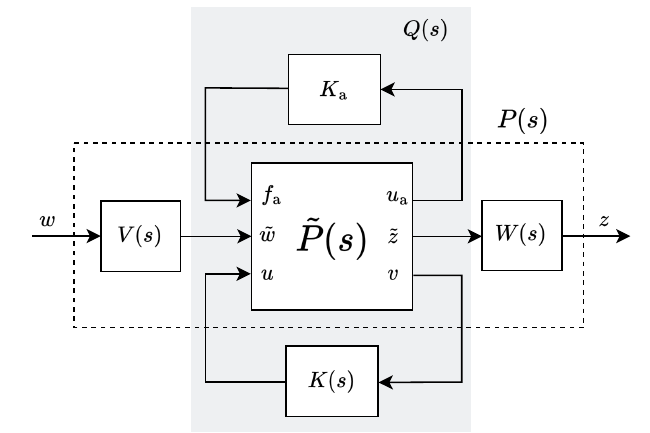}
    \caption{Generalized plant formulation of the control problem.}
    \label{fig:generalized_plant}
\end{figure}

Consider the generalized plant formulation of Figure \ref{fig:generalized_plant}. The generalized plant $\tilde{P}(s)$ is the open-loop mapping from disturbances $\tilde{w}$, controller command $u$, and actuator output $f_{\rm a}$ to the performance outputs $\bm{\tilde{z}}$, controller input $v$, and actuator input $u_{\rm{a}}$. The plant $P(s)$ includes the noise models and weighting filters. The controller $K(s)$ and actuator dynamics $K_{\rm a}$ are taken out of the loop because both are to be optimized by the algorithm. Since the disturbances do not have a unitary white power spectrum, we need to include noise models that colour white noise to the realistic disturbance spectra \cite{jabbenDEB}. The matrix $V(s)$ is a matrix with linear time-invariant noise colouring models on the diagonal. The individual entries colour white noise $w$ such that the coloured noise $\tilde{w}$ represents the actual disturbance and $\tilde{w} = V(s)w$. The entries of $V(s)$ thus consist of the DAC noise model, $V_{\rm{u}}(s)$, and the seismic disturbance model, $V_{\rm{x}}(s)$. The model $V_{\rm{x}}(s)$ is derived from known seismic disturbance spectra \cite{6d_isolation_system}, and includes the suspension dynamics, such that the seismic disturbance can be modelled as an additive noise source at the output of the plant. The matrix $W(s)$ is a diagonal matrix with designer-specified and potentially frequency-dependent weighting filters on the diagonal and combines the performance and robustness constraints.

\begin{align}
\left[\begin{array}{c}
\bm{u}_{\rm{a}} \\
\bm{\tilde{z}} \\
\bm{v}
\end{array}\right]=\tilde{P}(s) \cdot\left[\begin{array}{c}
\bm{f_{\rm{a}}} \\
\bm{\tilde{w}} \\
\bm{u}
\end{array}\right], \quad \tilde{\bm{w}}=\left[\begin{array}{l}
x_{\rm{d}} \\
u_{\rm{d}}
\end{array}\right], \quad \tilde{\bm{z}}=\left[\begin{array}{c}
\hat{x}\\
x \\
u \\
\end{array}\right],
\end{align}

\begin{equation}
    V(s) = \rm{diag}([V_{\rm{x}}, V_{\rm{u}}]), \hspace{5mm} W(s) = \rm{diag}([W_{\rm{p}},W_{\rm{dac}}, W_{\rm{rob}},W_{\rm{u}}]).
\end{equation}

The linear fractional transformation of the plant, actuators and controller is denoted by the transfer function $Q(s)$, equivalent to the closed-loop system that maps $\bm{\tilde{w}}\to \bm{\tilde{z}}$. The optimization problem is then defined by the following statement

\begin{equation}
\begin{aligned}
&\begin{aligned}
&\begin{aligned}
& \boldsymbol{K(s,K_{\rm{a}})}=\underset{K,\hspace{1mm}K_{\rm{a}}}{\arg \min }\left\|\left[\begin{array}{llll}
 \boldsymbol{W}_{\rm{p}} & & & \\
& \mathbf{0} & & \\
& & \mathbf{0} &\\
& & & \mathbf{0}\\
\end{array}\right] \boldsymbol{Q}(s)\left[\begin{array}{ll}
 \boldsymbol{V}_{\rm{x}}& \\
& \mathbf{0}
\end{array}\right]\right\|_2, \\
& \hspace{23.7mm}\text { s.t. }\left\|\left[\begin{array}{llll}
\mathbf{0} & & &\\
& \boldsymbol{W}_{\rm{dac}} & & \\
& & \mathbf{0} & \\
& & & \mathbf{0}
\end{array}\right] \boldsymbol{Q}(s)\left[\begin{array}{ll}
\mathbf{0} & \\
&\boldsymbol{V}_{\rm{u}} 
\end{array}\right]\right\|_{\infty}<\gamma_{\rm{d}}, \\
&\hspace{30.1mm}\left\|\left[\begin{array}{llll}
\mathbf{0} & & &\\
&  \mathbf{0} & & \\
& &\boldsymbol{W}_{\rm{rob}} & \\
& & & \mathbf{0}
\end{array}\right] \boldsymbol{Q}(s)\left[\begin{array}{ll}
\mathbf{1} & \\
& \mathbf{0}
\end{array}\right]\right\|_{\infty}<\gamma_{\rm{r}}, \\
&\hspace{30.1mm} \left\|\left[\begin{array}{llll}
\mathbf{0} & & &\\
&  \mathbf{0} & & \\
& &  \mathbf{0}&\\
& & & \boldsymbol{W}_{\rm{u,i}}
\end{array}\right] \boldsymbol{Q}(s)\left[\begin{array}{ll}
\boldsymbol{V_{\rm{x}}} & \\
& \boldsymbol{V}_{\rm{u}}
\end{array}\right]\right\|_2 \leq \gamma_{\rm{i}} \hspace{3mm}\forall \hspace{1mm} i, \\
& \hspace{30mm} K_{\rm{{lower, j}}} \leq K_{\rm{a, j}} \leq K_{\rm{upper, j}} \hspace{3mm} \forall \hspace{1mm} j.
\end{aligned}\\
\end{aligned}\\
\end{aligned}
\label{eq:optimization_1}
\end{equation}

\noindent where the actuator gains can be tuned by the optimizer, bounded by lower limit $K_{\rm{lower}}$ and upper limit $K_{\rm{upper}}$. The first $\mathcal{H}_{\infty}$-constraint is a limit on the open-loop equivalent spectrum of the DAC noise. The second $\mathcal{H}_{\infty}$-constraint determines the robustness, and the last $\mathcal{H}_2$-constraint ensures that each of the the actuator outputs remain with the DAC range. Typically, the cost function that is associated with such a mixed $\mathcal{H}_2/\mathcal{H}_{\infty}$ optimization problem is no longer convex. Although there exist methods to convexify this optimization problem \cite{scherer_mixed}, these can lead to conservative controller design, hence a non-smooth optimization algorithm  \cite{apkarian-2009, apkarian-2005} is utilized to solve for a controller $K(s)$ as well as the optimal actuator distribution $K_{\rm{a}}$. This optimization algorithm is implemented in \verb+Matlab's systune()+ function of the \verb+Control System Toolbox+.

\section{Results}

\begin{figure}[H]
    \centering
    \includegraphics[width=0.9\linewidth]{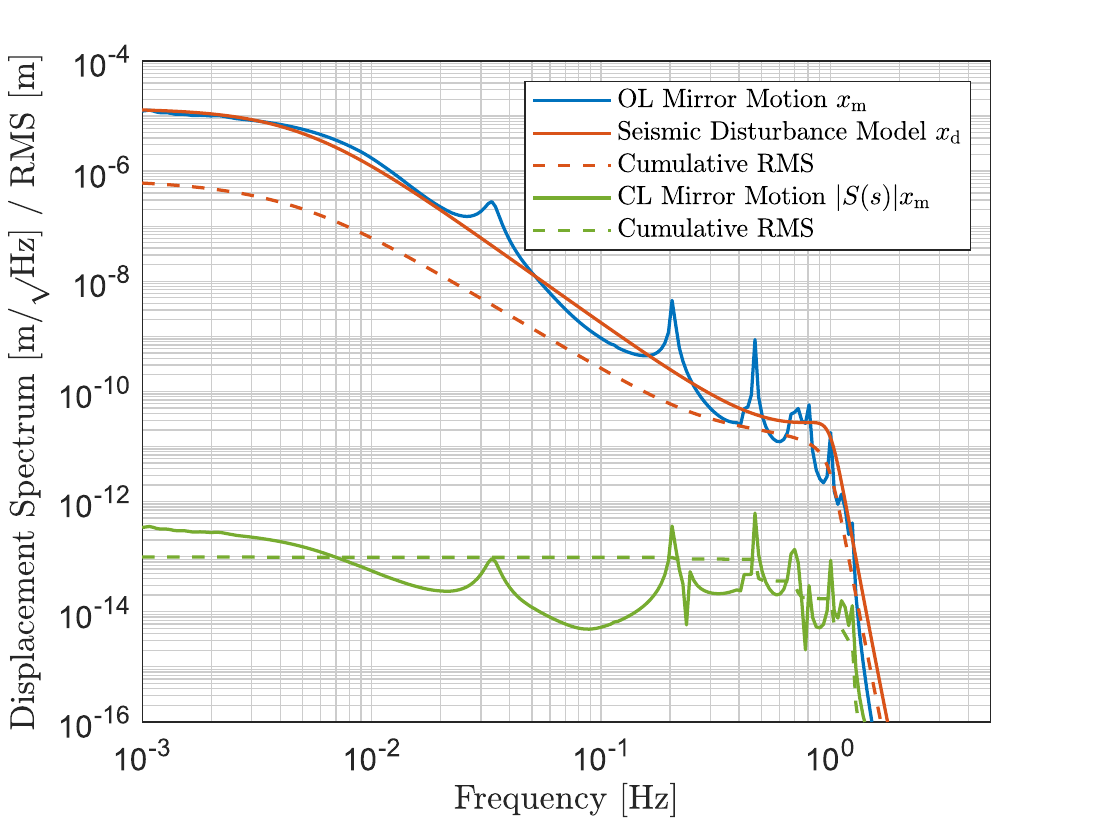}
    \caption{Closed-loop seismic disturbance compared with the open-loop. The simplified model $V_{\rm x}$ is used in the optimiser but performance is computed using the detailed disturbance.}
    \label{fig:cl_rms_seismic}
\end{figure}

The results of a controller optimization for the suspension model that was described in Section \ref{ch:modelling} are expressed in the open-loop DAC noise equivalence requirement and the suppression of the seismic disturbance. Figure \ref{fig:cl_rms_seismic} shows the cumulative $x_{\rm{rms}}$ of the open-loop motion, of which the lowest-frequency value corresponds to $x_{\rm{rms}}$. The cumulative RMS value is computed by integrating the spectra from high to low frequencies. The approximate model $x_{\rm{d}}$ is included. This simplified model is matched in RMS with the actual disturbance seen by the main observatory laser, $x_{\rm{m}}$. The value of $x_{\rm{rms}}$ is reduced to a value of $9.88 \cdot 10^{-14}$\,m, such that the seismic disturbance suppression requirement is satisfied.

\begin{figure}[H]
    \centering
    \includegraphics[width=0.9 \linewidth]{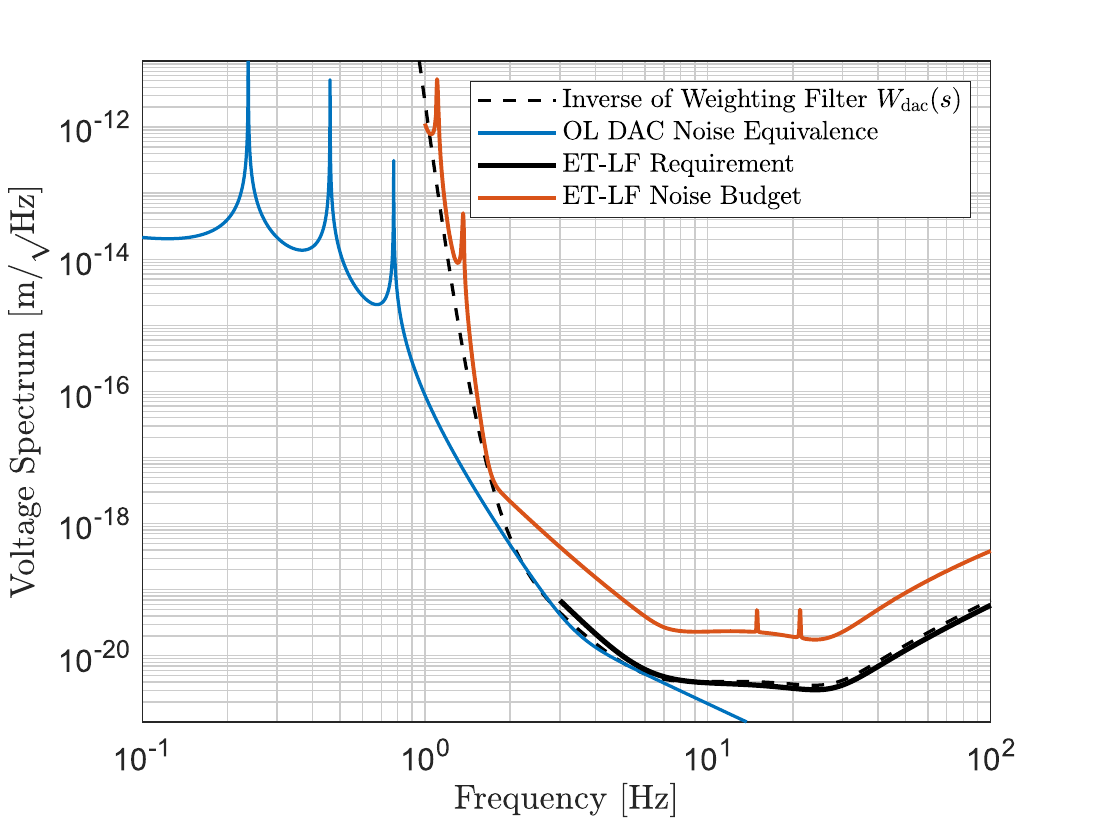}
    \caption{Closed-loop seismic disturbance suppression, compared to open-loop seismic disturbance}
    \label{fig:ol_dac}
\end{figure}

Figure \ref{fig:ol_dac} shows the open-loop equivalent DAC noise and the inverse of the weighting filter, $W_{\rm{dac}}^{-1}(s)$. The requirement includes a safety factor of 6 to keep the open-loop DAC noise well below the ET-LF sensitivity design curve. It can be seen that the open-loop DAC noise is tuned such that it touches the inverse of the weighting filter at around 2-5\,Hz, with $\gamma_{\rm{d}}  = 1.07$, which is something that one can typically expect from an $\mathcal{H}_{\infty}$-optimal controller since the objective according to Equation \ref{eq:optimization_1} is to minimize the $\mathcal{H}_2$-channel, as long as the $\mathcal{H}_{\infty}$-limits allow this.

\begin{figure}[H]
    \centering
    \includegraphics[width = 14cm]{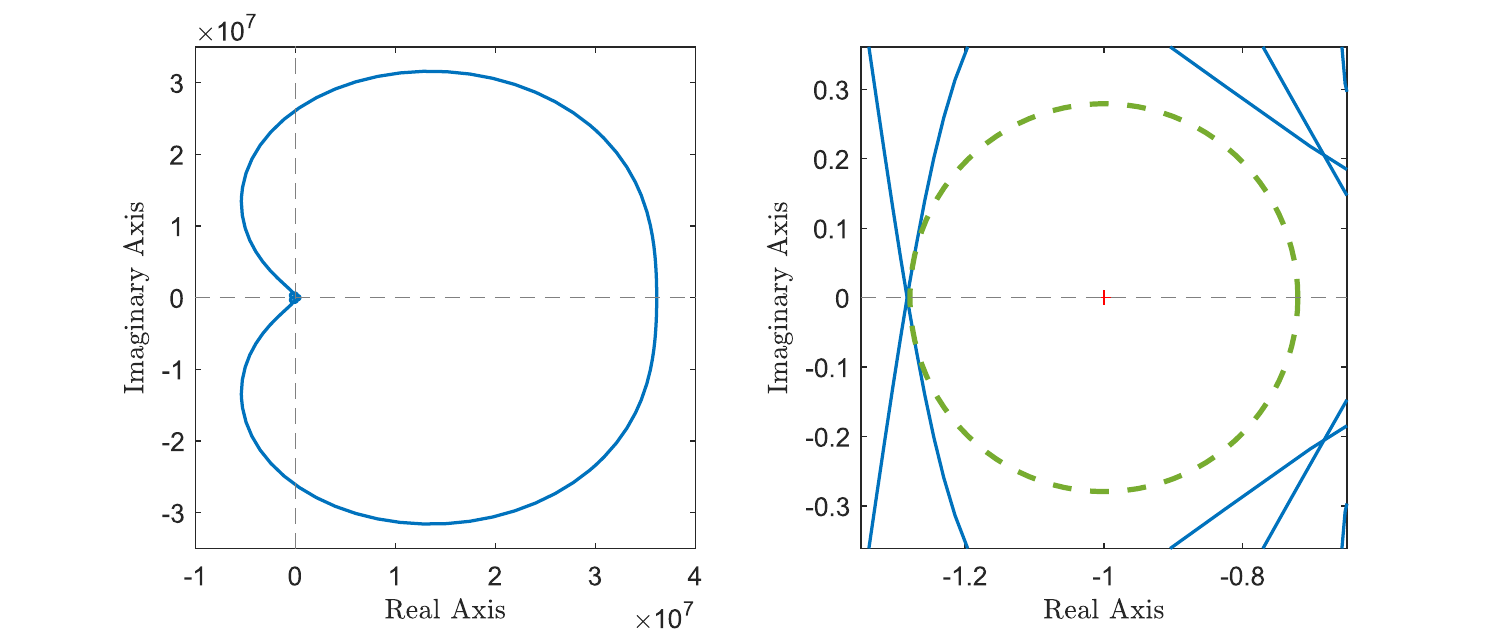}
    \caption{Nyquist plot of the loop gain $L(s)$ (left) and zoomed view around the critical point (right)}
    \label{fig:nyquist_plot}
\end{figure}

Additionally, the robustness constraint allows to obtain a desired modulus margin to guarantee sufficient margin against process variations. Usually, the dynamics of the suspension are well known for gravitational wave applications and the operating conditions are assumed to be rather constant. Therefore, it is not required to achieve large stability margins. The stability of the system is assessed via Nyquist stability requirement and the corresponding Nyquist diagram for the loop gain $L(s) = G(s)K_{\rm{a}}K(s)$ is shown in Figure \ref{fig:nyquist_plot}.

From the Nyquist plot, it can be concluded that the closed-loop system $Q(s)$ is stable and a requested modulus margin of 0.25 is guaranteed, which can be deduced from the green circle that indicates the area for which the distance to the critical point is less than 0.25. From this plot, it can be seen that the sensitivity $S(s)$ is tuned such that this closed-loop transfer function coincides exactly with $W_{\rm{rob}}(s)$ at multiple instances, reflected by a value of $\gamma_{\rm{r}} = 0.89$. Since the sensitivity touches the inverse robustness weighting filter at about $2-8\,\rm{Hz}$, the suppression of the low frequency seismic disturbance should be minimally affected by increased robustness margins.

\section{Discussion}

This study demonstrates a method for situating the control requirements, typical for a gravitational-wave observatory's test-mass suspension, within a non-convex, mixed optimization problem. The result of this study demonstrates that this methodology produces a quantitative global control distribution that satisfies the basic control requirements of the test-mass suspensions. Additionally, the optimizer allows to define open-loop weighting channels. For nominal $\mathcal{H}_2$ or $\mathcal{H}_{\infty}$ synthesis, it is often not possible to directly define such an open-loop requirement.

The controller is able to suppress the seismic noise sufficiently, such that the detectors' sensitivity is not degraded by the expense of excessive effects of the seismic noise at the mirror stage. From the open-loop DAC plot from Figure \ref{fig:ol_dac}, it can be seen that the graph closely meets the inverse of the weighting filter, which means that the controller is optimized such that the seismic motion is suppressed until the open-loop DAC noise injection limit is reached. Additionally, the weighting on the controller commands allows to ensure that the signal $u$ stays below the capacity of the actuators and drives. Finally, the Nyquist plot shows how an additional $\mathcal{H}_{\infty}$ robustness constraint allows to enforce a desired robustness margin such that the system is sufficiently far away from instability to account for process variations, such as process delays, which are inherently present to some extent in any control system.

A great benefit of optimization based control system design is that it alleviates the need for manual tuning of filters, allowing the global interferometric control distribution to be verified early in the design phase. Moreover, automated optimal control techniques allow to effectively assess many possible suspension configurations within the design space. It is often not possible to achieve this utilizing manual control design techniques and thus poses a great benefit in terms of allowable exploration of a more effective suspension design.

This specific case study involves for a simultaneously optimized actuator distribution for the suspension stages, however there are many more possibilities to study. The optimizer allows to explicitly tune parameterized models of dynamic systems, which allows for a simultaneous controller and suspension mechanics optimization. Since most of the time, extensive suspension models are present, this method lends itself well to integrate a suspension parameter optimization to fully benefit from the optimization strategy. This way, even better performance of the system may be achieved by having the optimizer aiding as a design tool for both controller algorithms and plant dynamics.

Finally, the controller that is tuned for this case study is a single transfer function matrix with three entries. Additionally, possible benefits of a multi degree-of-freedom controller may be studied, or the implementation of a feed-forward compensator might be of interest to include to further increase the performance of the system. All of these suggestions could be seamlessly integrated with the optimization study that was shown previously.


\section{Conclusions}

We showed that the combination of seismic disturbance variance minimization and frequency-dependent bounded requirements regarding the DAC noise can be condensed in a mixed $\mathcal{H}_2/\mathcal{H}_{\infty}$-optimization problem. Additional constraints, such as robustness and control signal variance constraints, can be integrated seamlessly with the optimization. Non-smooth synthesis was utilized to avoid conservative controller design. This method also explicitly allows for optimizing the actuation distribution and, moreover, lends itself well to include optimization of the mechanics or an additional feed-forward controller.
 

\vspace{6pt} 


\funding{This research received no external funding}

\dataavailability{The original data presented in the study are openly available in Zenodo; DOI 10.5281/zenodo.15126221}

\acknowledgments{The authors thank M. Valentini for useful discussions and input. This project has received funding from the European Research Council (ERC) under the European Union's Horizon 2020 research and innovation programme (grant agreement No. 865816).}

\conflictsofinterest{The authors declare no conflict of interest.} 

\newpage
\appendixstart
\appendix
\section[\appendixname~\thesection]{}
\subsection[\appendixname~\thesubsection]{}
\label{app:app1}
Table \ref{tab1} shows all control signals that appear in the figures and equations throughout this text. The following control signals can be identified. The argument $(s)$ is omitted for readability.

\begin{table}[H] 
\caption{Summary of the control signals, identified in the optimization problem.}
\label{tab1}
\begin{tabularx}{\textwidth}{LLL}
\toprule
\textbf{Signal}	& \textbf{Meaning} & \textbf{Unit}\\
\midrule

$f_{\rm a}$    & Actuator effort                     &   [N]\\
$u	$		    & Controller command voltage          & [V]\\
$u_{\rm a}$    & Actuator voltage, sum of $u$ and DAC noise & [V]\\
$u_{\rm d}$    & DAC noise & [V]\\
$v$             & Controller input & [m]\\
$w_1$           & White noise, coloured by $V_{\rm{u}}$  & [-]\\
$w_2$           & White noise, coloured by $V_{\rm{x}}$      & [-]\\
$w_{\rm rob}$  & Robustness constraint input, $w_{\rm{rob}} = x_{\rm d}$   &   [m]\\
$x$             & Total mirror motion      & [m]\\
$x_{\rm d}$    & Seismic disturbance felt at the mirror stage  & [m]\\
$x_{\rm m}$    & Real disturbance felt at the mirror stage             & [m]\\
$\hat{x}$       & Output of the suspension model             & [m]\\
$z_{\rm dac}$  & Open-loop DAC noise channel, $\mathcal{H}_{\infty}$-bounded  & [-]\\
$z_{\rm p}$    & Seismic disturbance suppression channel, $\mathcal{H}_2$-bounded & [-]\\
$z_{\rm rob}$  & Robustness channel, $\mathcal{H}_{\infty}$-bounded & [-]\\
$z_{\rm u}$    & Controller energy channel, $\mathcal{H}_2$-bounded & [-]\\

\bottomrule
\end{tabularx}
\end{table}

\subsection[\appendixname~\thesubsection]{}
\label{app:app2}
Table \ref{tab2} shows all control signals that appear in the figures and equations throughout this text. The following control signals can be identified. The argument $(s)$ is omitted for readability.

\begin{table}[H] 
\caption{Summary of the systems and models, identified in the optimization problem}
\label{tab2}
\begin{tabularx}{\textwidth}{LLLL}
\toprule
\textbf{Model}	& \textbf{Meaning} & \textbf{Unit}\\
\midrule

$G$             & Actuator effort                     &   [m/N]\\
$K$             & Controller                          &   [N/m]\\
$K_{\rm{a}}$    & Actuator dynamics                   &   [N/V]\\
$P$             & Generalized plant                   & [-]\\
$\tilde{P}$     & Generalized plant without weights and noise models  & [-]\\
$Q$             & Linear fractional transformation of $P$, $K$ and $K_a$ & [-]\\
$S$             & Sensitivity transfer function       & [-]\\
$V$             & Matrix with noise models on the diagonal & [-]\\
$V_{\rm{x}}$	& Seismic disturbance model           &   [m]\\
$V_{\rm{u}}$    & DAC noise model                     &   [V]\\
$W$             & Matrix with weighting filters on the diagonal & [-]\\
$W_{\rm{dac}}$  & $\mathcal{H}_{\infty}$-weight open-loop DAC noise      & [-]\\
$W_{\rm{p}}$    & $\mathcal{H}_2$-weight seismic disturbance      & [-]\\
$W_{\rm{rob}}$  & $\mathcal{H}_{\infty}$-weight, robustness filter                 & [-]\\
$W_{\rm{u}}$    & $\mathcal{H}_2$-weight controller command  & [-]\\

\bottomrule
\end{tabularx}
\end{table}
\newpage
\subsection[\appendixname~\thesubsection]{}
\label{app:app3}

The open-loop gain $L(s)$ resulting from the optimization is shown in Figure \ref{fig:loopgain}. The gain and phase margin can be computed as:

\begin{figure}[ht!]
    \centering
    \includegraphics[width=0.9\linewidth]{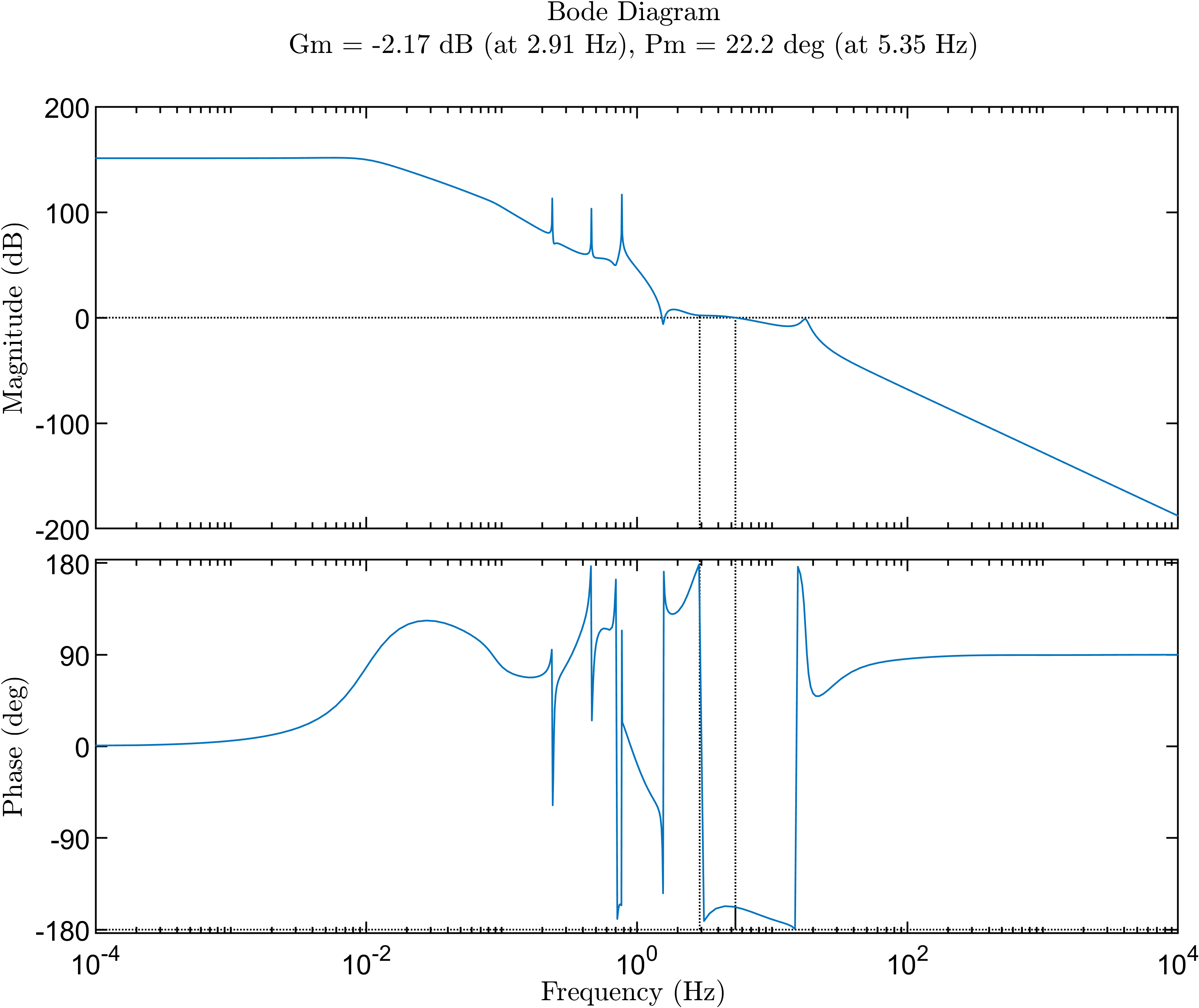}
    \caption{Bode plot of the loop gain $L(s) = G(s)K_{\rm a}K(s)$}
    \label{fig:loopgain}
\end{figure}

The sizing of the actuators is expressed in terms of achievable displacement in meters per volt for each actuator. The sizing is then computed as $K_{\rm a}G(0)$. The values that resulting from the optimization study are summarized in Table \ref{tab:tab3} below, as well as the RMS of the control commands for each channel.

\begin{table}[H] 
\caption{Sizing of the actuators, resulting from the optimization study}
\label{tab2}
\begin{tabularx}{\textwidth}{LLLL}
\toprule
&\textbf{Top actuator}	& \textbf{Middle Actuator} & \textbf{Bottom Actuator}\\
\midrule
$K_{\rm a}$ & $1.0\cdot10^{-7} [{\rm m/V}] $  & $1.2\cdot10^{-9}[{\rm m/V}]$ & $7.2\cdot10^{-11}[{\rm m/V}] $ \\
$u_{\rm rms}$ & $6.0 {\rm V}$ & $1.5 {\rm V}$ & $1.3 {\rm V}$\\

\bottomrule
\end{tabularx}
\label{tab:tab3}
\end{table}
\begin{adjustwidth}{-\extralength}{0cm}
\printendnotes[custom] 

\reftitle{References}

\PublishersNote{}
\end{adjustwidth}
\end{document}